\documentclass[12pt,preprint]{aastex}
\input{epsf}
\begin{document}

\title{Binary Star Orbits. III. In which we Revisit the \\ Remarkable Case 
of Tweedledum and Tweedledee}

\author{Brian D. Mason\altaffilmark{~1} and William I. 
Hartkopf\altaffilmark{~1}}
\affil{U.S. Naval Observatory \\ 3450 Massachusetts Avenue, NW, Washington,
DC, 20392-5420 \\ Electronic mail: (bdm, wih)@usno.navy.mil}

\author{Harold A. McAlister}
\affil{Center for High Angular Resolution Astronomy\\Georgia State 
University, Atlanta, Georgia  30303\\Electronic mail: hal@chara.gsu.edu}

\altaffiltext{1}{Visiting Astronomer, Kitt Peak National Observatory and 
Cerro Tololo Interamerican Observatories, National Optical Astronomy 
Observatories, operated by the Association of Universities for Research in 
Astronomy, Inc., under contract with the National Science Foundation.}

\begin{abstract}

Two of the most challenging objects for optical interferometry in the middle
of the last century were the close components (FIN 332) of the wide 
visual binary STF2375 (= WDS 18455$+$0530 = HIP 92027 = ADS 11640). Each 
component of the wide pair was found to have subcomponents of approximately 
the same magnitude, position angle and separation and, hence, were 
designated by the tongue in cheek monikers ``Tweedledum and Tweedledee''
by the great visual interferometrist William S. Finsen in 1953. They were 
later included in a list of ``Double Stars that Vex the Observer'' by W.H. 
van den Bos (1958a).

While speckle interferometry has reaped a rich harvest investigating the 
close inteferometric binaries of Finsen, the ``Tweedles'' have continued to
both fascinate and exasperate due to both the great similarity of the close 
pairs as well as the inherent 180$^{\circ}$ ambiguity associated with 
interferometry. 

Detailed analysis of all published observations of the system have revealed
several errors which are here corrected, allowing for determination of these
orbital elements which resolve the quadrant ambiguity. A unique software 
filter was developed which allowed subarrays from archival ICCD speckle data
from 1982 to be re-reduced. Those data, combined with new and unpublished 
observations obtained in 2001-9 from NOAO 4m telescopes, the Mt.\ Wilson 
100in telescope and the Naval Observatory Flagstaff Station 61in telescope 
as well as high quality unresolved measures all allow for the correct orbits
to be determined. Co-planarity of the multiple system is also investigated.

\end{abstract}


\section{The discovery and early measures of Tweedledum and Tweedledee}

The bright star HR 7048 [ = HD 173495 = HIP 92027 = ADS 11640 = STF2375, 
($\alpha$,$\delta$) = 
18$^{\rm h}$45$^{\rm m}$28$^{\rm s}$\llap.4 $+$05$^{\circ}$30$'$00$''$ 
(2000)] was first recognized as a double by F.G.W. Struve in 1825 (Struve 
1837). Since that time the system has been well observed by many double star
astronomers, and has probably been most useful for those wishing to 
characterize or calibrate their observational systematics, as the motion has 
long been recognized as quite slow. As early as the start of the last 
century, Burnham (1906) in his double star catalog (where this object is 
listed at \# 8776) noted  ``no change in distance, and but little, if any, 
in the angle.'' Almost a century later we have seen a cumulative change of 
only 12$^{\circ}$ and 0$''$\llap.3 since the discovery epoch. The first 
indication that this system might be more than just a slowly moving pair 
came in a compelling note to Aitken's (1932) catalog  which stated that the 
``radial velocity of A is variable with a range of 99 km/sec,'' citing no 
less an authority than Plaskett et al.\ (1921) as the source. The source of 
the variability --- and the system's interest --- was discovered by William 
Finsen some two decades later.

After experimenting with different interferometer designs, Finsen (1964a) 
had constructed an eyepiece interferometer, where the observer visually 
measured interferometric fringe visibilities, then calculated position 
angles and separations. This instrument, as with other interferometric 
techniques, was best suited to brighter stars and therefore, a program 
commenced to investigate the duplicity of all 8,117 stars brighter than
magnitude 6.5 with $+$20$^{\circ}$~$<$~$\delta$~$<$~$-$75$^{\circ}$. In 
addition to measuring many thousands of known systems, application of this 
new technique starting in 1951 (Finsen 1951) led to the discovery of 79 new 
pairs (Mason et al.\ 2001) almost all of which are close and astrophysically 
interesting due to their rapid motion.

However, upon turning to the wide components of the Struve pair, Finsen was 
initially surprised and confused. His first observations were rather vexing,
as he reported in his article {\it A case of Tweedledum and 
Tweedledee\footnote{Tweedledum and Tweedledee are nursery rhyme characters 
whose names first appeared in an epigram by John Byrom (1692--1763). They 
are best known as a pair of identical twins reciting these rhymes in Lewis 
Carroll's {\it Through the Looking Glass and what Alice Found There} 
(Dodgson 1871).}} (Finsen 1953, 1954):

\begin{quote}

``The two pairs are remarkably similar; in fact the simultaneous 
disappearance of both sets of fringes gave rise to considerable dismay till 
careful checking showed there was nothing wrong with the instrument.''

\end{quote}

Apparently, Finsen was quite careful and did not trust his result fully 
until independently confirmed and measured (albeit crudely) with a 
micrometer by van den Bos (1956) two nights later. These measures were also 
quoted in Finsen's discovery paper (1953, 1954). The systems were observed 
and reported on a fairly regular basis in the 1950s and early 1960s, with 
eyepiece interferometry by Finsen and with micrometry on large refractors 
and reflectors by van den Bos, van Biesbroeck, and Muller (see Tables 1 and 
2 for all measures and references). Both systems then disappeared from sight
(like the Cheshire Cat?) as reported by Finsen (1965, 1967, 1969), van den 
Bos (1963b), and Worley (1972), although Walker (1969) listed a measure for 
Ba,Bb (then designated CD) obtained in 1966. 

Throughout this time STF2375 retained considerable interest and was among 
six systems described in some detail by van den Bos (1958a) in his article 
{\it Double Stars that Vex the Observer}, where he elaborated a bit upon 
Finsen's discovery:

\begin{quote}

``\ldots When inspecting ADS 11640, Finsen was startled to see the fringes 
on both components of the Struve pair disappear simultaneously when rotating
the interferometer. He suspected, at first, that something had gone wrong 
with the instrument, but other stars showed nothing abnormal and it turned 
out that he had indeed found, not fraternal but identical twins, for which 
he applied the nicknames `Tweedledum and Tweedledee.'

I have recently measured this object with the Lick 36-inch refractor which 
clearly separates the two close pairs and the appearance is astonishing. 
Apart from the fact that Tweedledee \ldots is slightly fainter than 
Tweedledum \ldots, I can see no difference between the two \ldots ''

\end{quote}

\section{Getting too close to resolve}

From the first work of John Herschel (1847), through the large survey of 
Rossiter (1955) and the work of Finsen and van den Bos at Union and (later) 
Republic Observatory, double star work at the Cape could be characterized in
one of two ways: excellent or inactive. The disappearance of the Tweedles in
retrospect seemed to portend a period of benign neglect at the Cape, as van 
den Bos left for the United States and Finsen approached retirement with 
some trepidation, as he wrote to Charles Worley (1968a):

\begin{quote}

``\ldots There is a move afoot to give first and absolute priority to a 
programme of planetary photography, to the distress of van den Bos and 
myself \ldots and this seems likely to ring the death knell of our long 
record of intensive double star observing. I found it impossible to explain 
to people with no experience of double star observing of the demands it 
places on the observer's skill, enthusiasm and energy to relegate that to a 
second priority time-filling role is very discouraging, to say the least of 
it, and may very well kill it stone dead. Time will show.''

\end{quote}

As Finsen predicted, double star astronomy in South Africa saw a definite 
downturn after his retirement. Fortunately, some ten years after the demise 
of eyepiece interferometry the technique of speckle interferometry was 
developed. In the late 1970s one of the authors (H.A.M.) began a healthy 
correspondence with W.S. Finsen just as his speckle program was getting 
started, regarding objects which would be suitable for speckle 
interferometry. Finsen's continued interest in this pair is apparent in his 
letter of 1977:

\begin{quote}

``\ldots I was reminded of the quadruple that I have dubbed `Tweedledum and
Tweedledee' \ldots Have you got this on your programme? It would be fun if 
you could follow it up and eventually do the orbits. These `identical twins'
caused me much agony of mind before I was prepared to accept their duplicity
as real. I measured them regularly until 1963 when both became too close to 
measure without much change in position angles.''

\end{quote}

\section{Speckle Interferometry: The reappearance of the Tweedles}

Due to their spatially close nature, many of the systems first resolved by 
Finsen have orbital periods of less than 50 years. Thus speckle 
interferometry became a mature technique at an optimal time for orbital 
analysis of many of Finsen's discoveries; observation of the Finsen stars 
was therefore given high priority in the early years of this technique. 
Early results of those efforts include orbital analyses of FIN 342 
(McAlister et al.\ 1988), FIN 312 (Hartkopf et al.\ 1989), FIN 331, 325, 
350, 381 (Hartkopf et al.\ 1996), FIN 347 (Mason et al.\ 1996), FIN 359 
(Mason 1997), FIN 47 and 328 (Mason et al.\ 1999). 

FIN 332 Aa,Ab and Ba,Bb were both recovered by speckle interferometry in 
1976, and continued to be observed on numerous occasions by this and other 
interferometric techniques (see Tables 1 and 2 and reference quoted 
therein). 

Figure 1 presents a demonstration of the similarity in spatial 
characteristics of the Tweedles in a ``Ferris Wheel'' plot. In this diagram 
the two pairs are shown relative to each other and to the same scale and 
orientation. The small ellipse in the lower left is the calculated orbit of 
FIN 332 Aa,Ab while the one in the upper right is Ba,Bb. The large dashed 
ellipse is an indication of the motion of the wider pair, although given the
very small coverage of the orbit it is only present to give an idea of the 
relative scales of the orbits. The axes are in seconds of arc. The orbits of
the close pairs are described in \S 5.1. 

\placefigure{tweedle_fig1}

\begin{figure}[!ht]
\epsfxsize 5.0in
\centerline{\epsffile{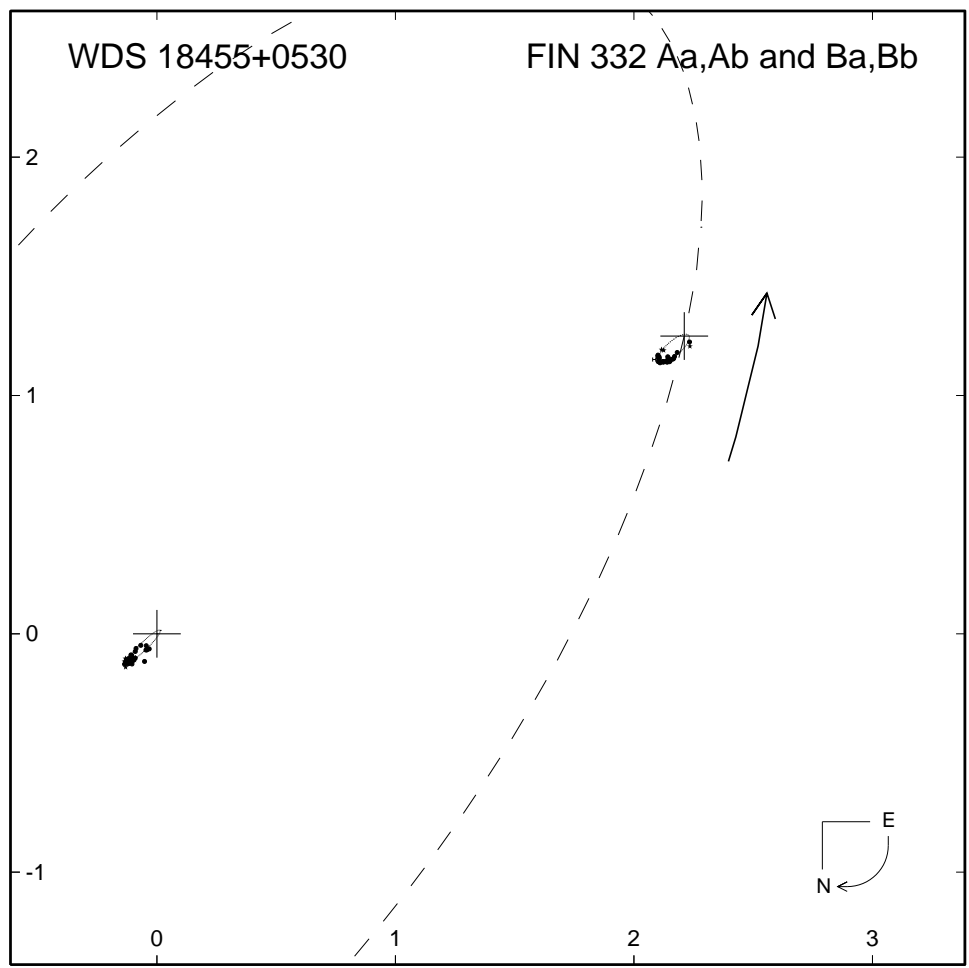}}
\caption{A ``Ferris Wheel'' plot of FIN 332 Aa,Ab and Ba,Bb, shown to the 
same scale as that of the wide pair, STF2375. These are the orbits of 
Figures 5 and 6. In this plot the relative positions of STF2375 A and B are 
fixed and the dashed curve is indicative of the pair's orbital motion 
(although it has only moved 12$^{\circ}$ since its discovery in 1825, so no 
believeable orbital elements can be determined). The amount and direction of
motion of the AB pair over the past 185 years are indicated by the thick 
curved arrow. The arrow at lower right indicates the direction of motion of 
both close pairs, which is opposite that of the wide pair. Scales are in 
arcseconds.}
\label{fig1}
\end{figure}

\section{Measures of the Tweedles}

Tables 1 and 2 present the observations of FIN 332 Aa,Ab and Ba,Bb 
respectively. Columns one through four contain data specific to the 
observation: the epoch of observation (expressed as fractional Besselian 
year), position angle (in degrees), separation (in seconds of arc), and 
number of measures comprising this mean published position. Note that the 
position angle has not been corrected for precession and is thus based on 
the equinox for the epoch of observation. When the pair is unresolved the 
lower limit on separation is given in column three if published or 
determined here. Columns five and six give O$-$C orbit residuals (in 
$\theta$ and $\rho$) to the orbits presented in \S 5.1. When the components 
are unresolved, the O$-$C columns (five and six) now give, in parentheses, 
the position predicted by these orbits. The method of observation is 
indicated in column seven, while the reference to the measure is in column 
eight. The Center for High Angular Resolution Astronomy (CHARA) photographic
speckle camera was less sensitive than the ICCD system, as seen by the small
number of measures of the fainter Ba,Bb pair (N=8) vs. Aa,Ab (N=17). 
Finally, column nine is reserved for the many notes to the measures. In 
addition to quadrant flips indicated by the correct determination of this 
previously ambiguous characteristic, there are also other cases where the 
originally published measures have been corrected. These are described in \S
4.1.

A representation of the similarity of measurements of these systems to each 
other is presented in Figures 2a and 2b. Note that the predicted separation 
and position angle differences (assuming an arbitrary quadrant, i.e., 
$\pm$180$^{\circ}$) are usually quite small, especially at the time of the 
discovery and during the first phase of resolutions (\S 1) where 
$\Delta \rho <$ 0\farcs04 and $\Delta \theta <$ 3$^{\circ}$. The two curves 
and shaded regions, representing the orbital solutions, are presented below.

\placefigure{tweedle_fig2a,tweedle_fig2b}
 
\begin{figure}[t]
\centerline{\epsfxsize 3.5in \epsffile{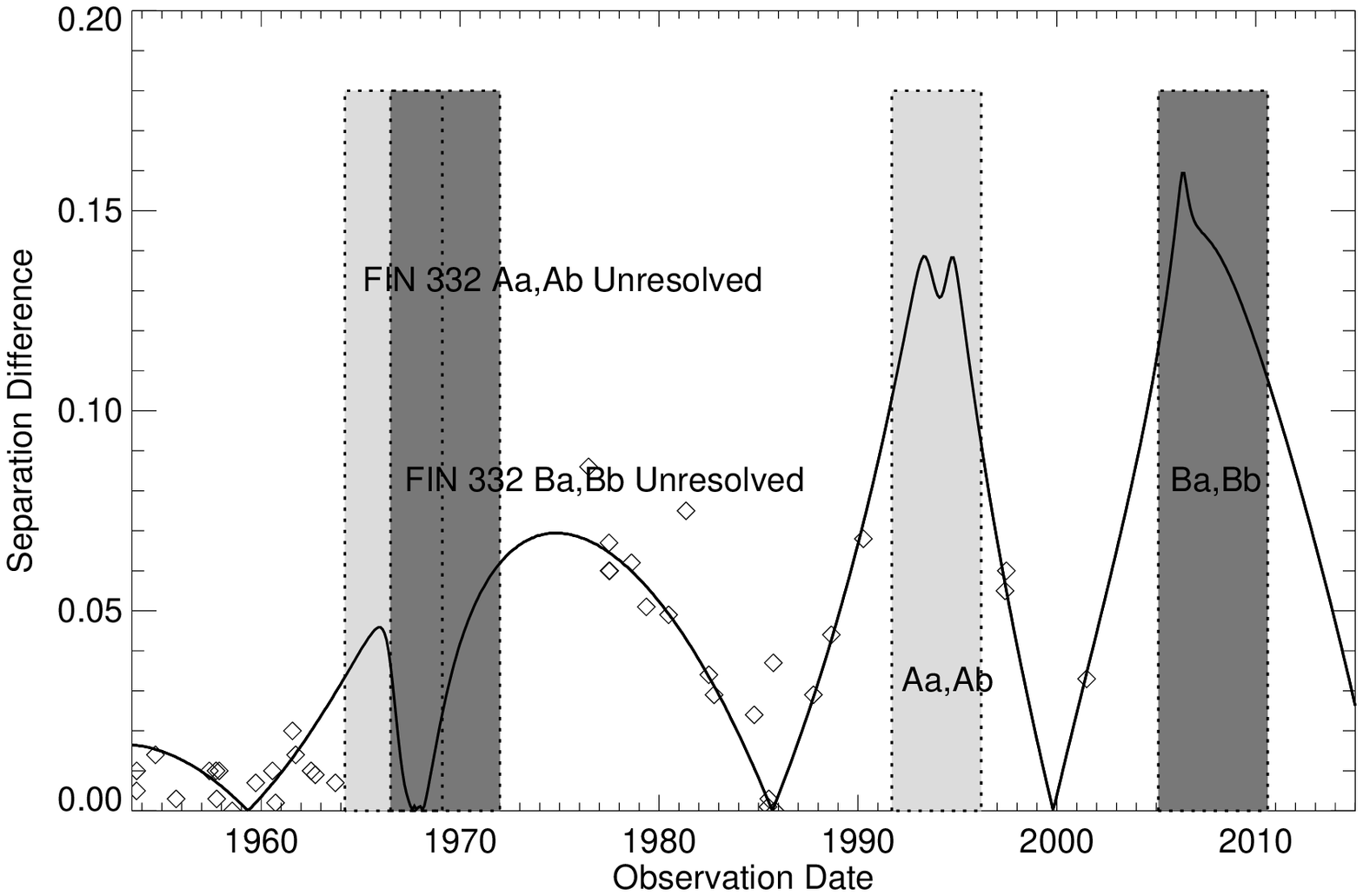} \epsfxsize 3.5in \epsffile{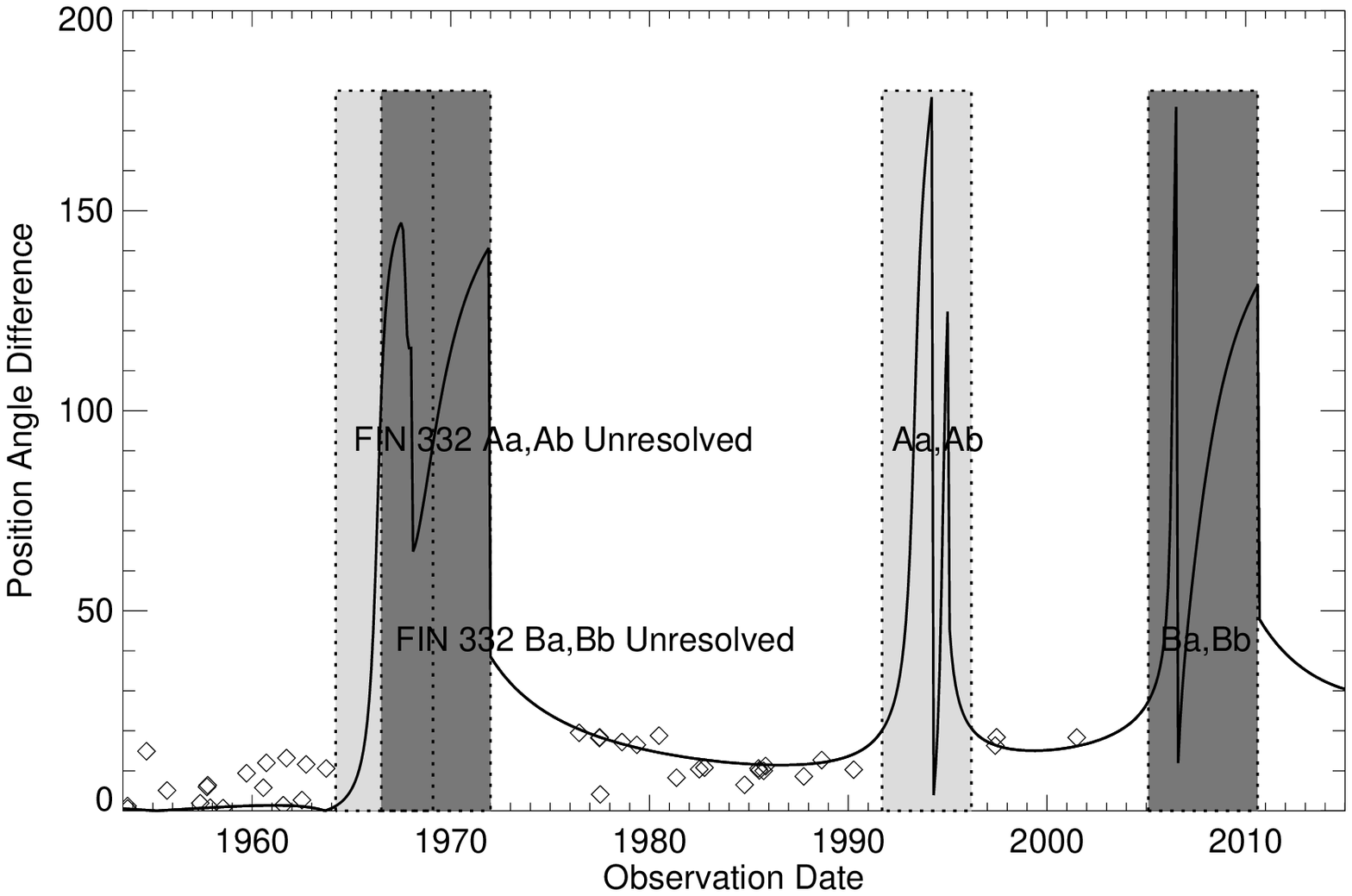}}
\caption{The difference in separation and position angle between FIN 332 
Aa,Ab and Ba,Bb. The solid curve indicates the predicted difference based on
the orbits of \S 5.1. The diamonds represent the differences between 
measures of Aa,Ab and Ba,Bb when they were made at the same time by the same
observer. The lighter shaded areas from 1964.2--1969.1 and 1991.7--1996.2 
are dates when Aa,Ab is predicted to be closer than 0\farcs05. The darker 
shaded areas from 1966.5--1972.0 and 2005.1--2010.6 indicates the range of 
dates when the orbit predicts Ba,Bb to be closer than 0\farcs05. }
\label{fig2}
\end{figure}

Table 3 provides measures, contemporaneous with those new measures presented
here, of the wider AB parent pair, STF2375. Columns one through four are as
Tables 1 and 2. Column five provides the method while Column 6 the notes. In
this case, the notes simply indicate which telescope was used.

\subsection{Corrections to Published Measures}

A total of seven measures (Aa,Ab: 4, Ba,Bb: 3) were initially published by 
CHARA using a preliminary calibration. This calibration was corrected in 
McAlister et al.\ (1989) and the corrected measures first appeared in 
McAlister \& Hartkopf (1988); the corrected measures are listed here. While 
the very small $\Delta$m of Aa relative to Ab and of Ba relative to Bb 
presents one set of unique problems, another is the small (but easily 
measurable) difference between the different components in the wide STF2375 
system. Normally, the field-of-view is such that it is possible to observe 
both pairs; however, problems with analysis of the complex system (see \S 
4.4 below), led the CHARA collaboration to observe this system under high 
magnification so that only one pair was resolved at a time. In this case, a
measure of the Ba,Bb pair (Hartkopf et al.\ 1997) was incorrectly assigned 
to Aa,Ab.

Among the most difficult sets of observations to sort out were the 1984 and 
1985 observations of the Ba,Bb pair made by Tokovinin \& Ismailov (1988). 
After investigating numerous possible quadrant flips and/or identification 
errors for these measures and incorporating the unresolved measures in the 
analysis the two measures did not fit any orbital analysis obeying Kepler's
Laws. The first author was consulted, and it is possibly most instructive 
at this point to quote directly from his response (Tokovinin 2001):

\begin{quote}

``I observed it myself at the 1m telescope in 1981, 1984, 1985 with the 
phase grating interferometer. Both close pairs fell within the focal 
aperture, so for this object I had to de-center and hide either Aa,Ab or 
Ba,Bb (in your notation) behind the diaphragm, to get the visibilities of 
the remaining pair. It is very unlikely that I misidentified the close 
pairs... For pairs of this separation, the curve of visibility vs. $\theta$ 
has two rather similar maxima. Apparently, in reducing the 1984--85 Ba,Bb 
data I took the wrong one: this changes the P.A. by roughly 90$^{\circ}$, 
and gives similar, but wrong separation. The choice of the `correct' maximum
was often guided by the previous measurements, and,  apparently, in this 
case was wrong! So, the data on Aa,Ab as measured in 1984--85 must be still 
valid, and not attributed to Ba,Bb. Measurement error, however, can still be
too big, compared to the 4m speckle, because it's a difficult object, it was
de-centered, etc. \ldots''

\end{quote}

Given this, it is not surprising, despite the 90$^{\circ}$ adjustment, that 
these measures had residuals judged too large by the orbit calculation.

\subsection{Hipparcos}

The Hipparcos satellite (ESA 1997) observed STF2375 and resolved the wider 
AB pair and the Ba,Bb pair at the calculated date of 1991.25. The Aa,Ab pair
was not resolved. Due to the substantial orbital motion of the Ba,Bb pair 
during the course of the Hipparcos mission, the quality of this measure may 
be somewhat degraded.

While all components have the same parallax and proper motions [$\pi$ = 
4.60$\pm$1.10 milliarcseconds (mas), $\mu_{\alpha}$ = 15.54$\pm$1.07 mas, 
$\mu_{\delta}$ = 1.96$\pm$0.86 mas (ESA 1997) $\pi$ = 5.30$\pm$0.85 
milliarcseconds (mas), $\mu_{\alpha}$ = 14.32$\pm$1.04 mas, $\mu_{\delta}$ =
0.16$\pm$0.87 mas (van Leeuwen 2007)], the errors are probably larger than 
this, possibly by as much as 50\%  (see Urban et al.\ 2000). While the 
reasons for the error underestimation may be complex, it is likely that long
term motion of wide pairs may not be fully characterized in the few year 
Hipparcos solution. The Tycho-2 (H$\o$g et al.\ 2000a,b) proper motion, 
which is possibly more accurate for long period doubles like AB, is 
$\mu_{\alpha}$ = 9.4$\pm$2.0 mas, $\mu_{\delta}$ = $-$2.5$\pm$1.9 mas. 
Normally a system this bright would have many historical measures to improve
the proper motion. However, as the AB system was judged a close pair it was 
left off many transit circle programs; and only three historical measures, 
Albany 10, AGK2 and AGK3, were used in the Tycho-2 proper motion 
determination; all of these were photocenter observations (Urban 2002). Of 
greater concern is the large $\Delta$m assigned by Hipparcos to the Ba,Bb 
pair of H$_{\rm p}$ = 0.76 mag. Although the errors are large (Ba = 
7.192$\pm$0.155 mag, Bb = 7.952$\pm$0.315 mag), it is certainly difficult to
reconcile this large $\Delta$m with the many visual estimates. It may be the
result of too many free parameters for four physically related components, 
even though only three were resolved. 

\subsection{The Magnitude Difference}

Based on published pre-speckle magnitude difference estimates, Aa,Ab has a 
mean $\Delta$m of 0.008$\pm$0.119 while Ba,Bb has a mean of 0.043$\pm$0.232.
This quadrant ambiguity can result in two consistent results: one solution 
is of period $P$ and high eccentricity and, contrariwise (as one of the 
Tweedles might say), another solution is of period $2P$ and low 
eccentricity. While we have contemporary measures of $\Delta$m (see the 
notes to Table 1 \& 2) which are larger, this only gives the absolute 
orientation at a single epoch; establishing which orbit is correct requires
information from data at either end of the long period solution. The 
quadrant analysis of Bagnuolo et al.\ (1992) was successfully used on FIN 
342 (McAlister et al.\ 1988), another binary of small $\Delta$m, and this 
method was utilized here to definitely establish the correct quadrant for 
both pairs using both historical CHARA ICCD speckle data and more recent 
United States Naval Observatory (USNO) ICCD speckle data. While preliminary 
analysis (Mason \& Hartkopf 2002) generated long- and short-period solutions
for both close pairs, the short-period, high-eccentricity solution has now 
been determined to be correct in both cases.

\subsection{New Old Measures}

The first two measures taken with the CHARA CCD system were obtained at a 
relatively low magnification, such that both of the wider components were 
observed in the same dataset. As a result of their similar morphologies the 
closer subcomponents were found to overlap in Fourier space. Of the thirteen
peaks ($n(n-1)+1$) expected to be seen in autocorrelation space for a 
quadruple, only nine were seen. Figure 3a is the measured system geometry at
the time of this observation and Figure 3b illustrates the resulting 
autocorrelogram. Figure 3c is the actual ``full frame'' directed vector 
autocorrelation (DVA) of the 1982 data. 

\placefigure{tweedle_fig3a,tweedle_fig3b,tweedle_fig3c}
 
\begin{figure}[!ht]
\centerline{\epsfxsize 2.5in \epsffile{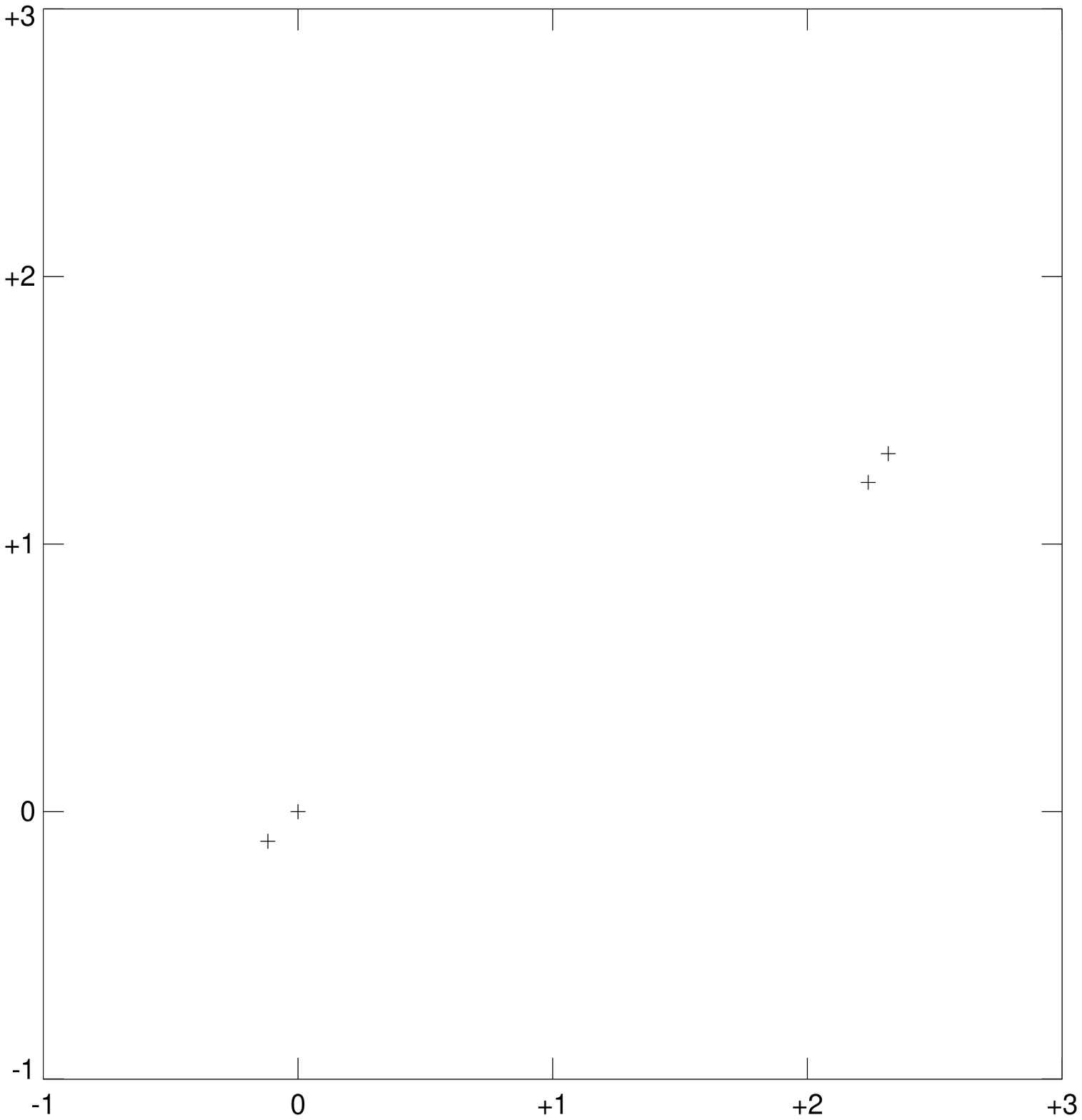} \epsfxsize 2.5in \epsffile{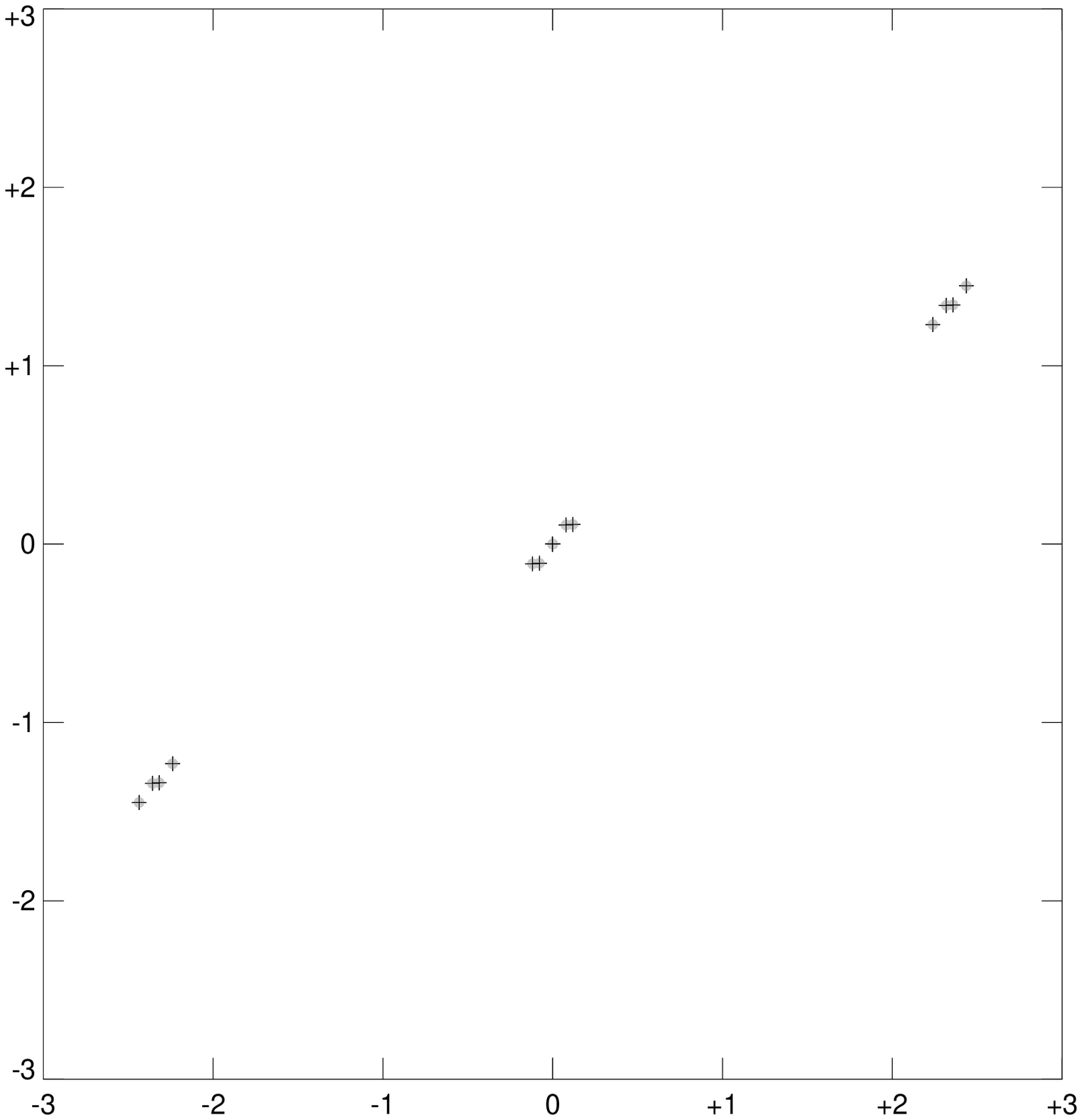} \epsfxsize 2.4in \epsffile{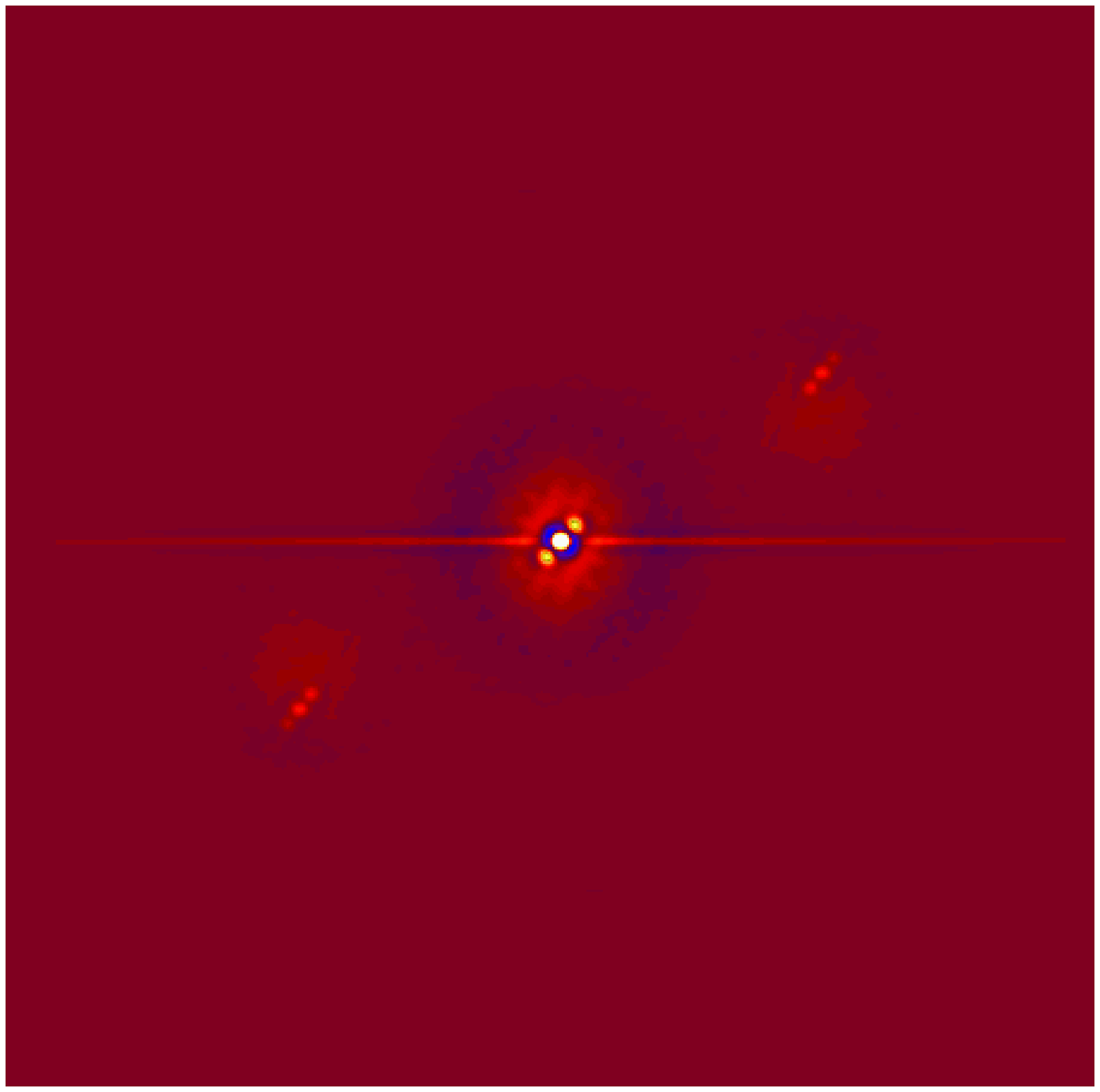}}
\caption{Panel a (left) is a representation of the geometry of the system at
1982.7650 based on the new measures of the 1982 data. The scales are in 
seconds of arc. The origin is the location of the Aa component in relative 
astrometry space. Panel b (center) is the autocorrelation of a. Note the 
four visible ``double $+$ signs'' which represent the blends described in 
\S 4.4. The central peak is the zeroth order autocorrelation spike. The gray
circles, barely visible at this scale, are 0\farcs030 in diameter to 
indicate regions where detail cannot be seen due to the resolution 
capabilities of the telescope. Panel c (right) is a digitization of the 
1982.7650 data with the blended images. While some of the peaks are quite 
faint, all nine visible peaks of b are seen here.}
\label{fig3}
\end{figure}

In mid-2007, Ellis Holdenried\footnote{USNO, retired.} developed software 
for calculating the DVA of a user-defined subarray of a CCD frame. Review of
the archived videotape data, obtained in 1982 and dubbed in 1995, seemed to 
indicate that the tracking of the telescope was adequate and seeing was 
good, such that the selected subarray could be static rather than dynamic. 
While there was some degradation in the video signal, there is significant 
past experience in working with these old data and recovering good science 
(Hartkopf et al.\ 2000). Results of the application of the Holdenried 
subarray DVA for the two pairs are illustrated in Figure 4.

\placefigure{tweedle_fig4a,tweedle_fig4b}
 
\begin{figure}[!ht]
\centerline{\epsfxsize 3.5in \epsffile{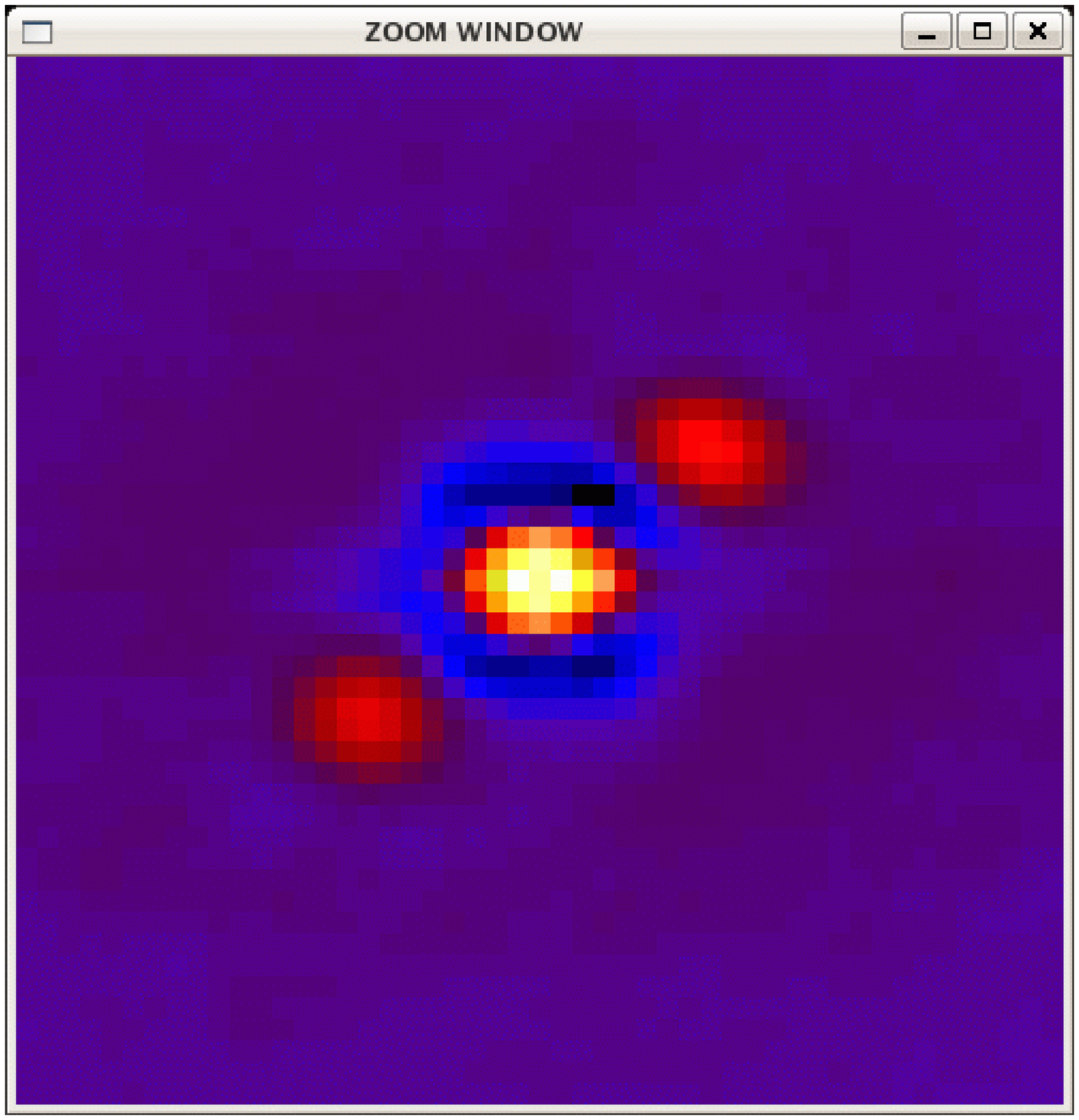} \epsfxsize 3.5in 
\epsffile{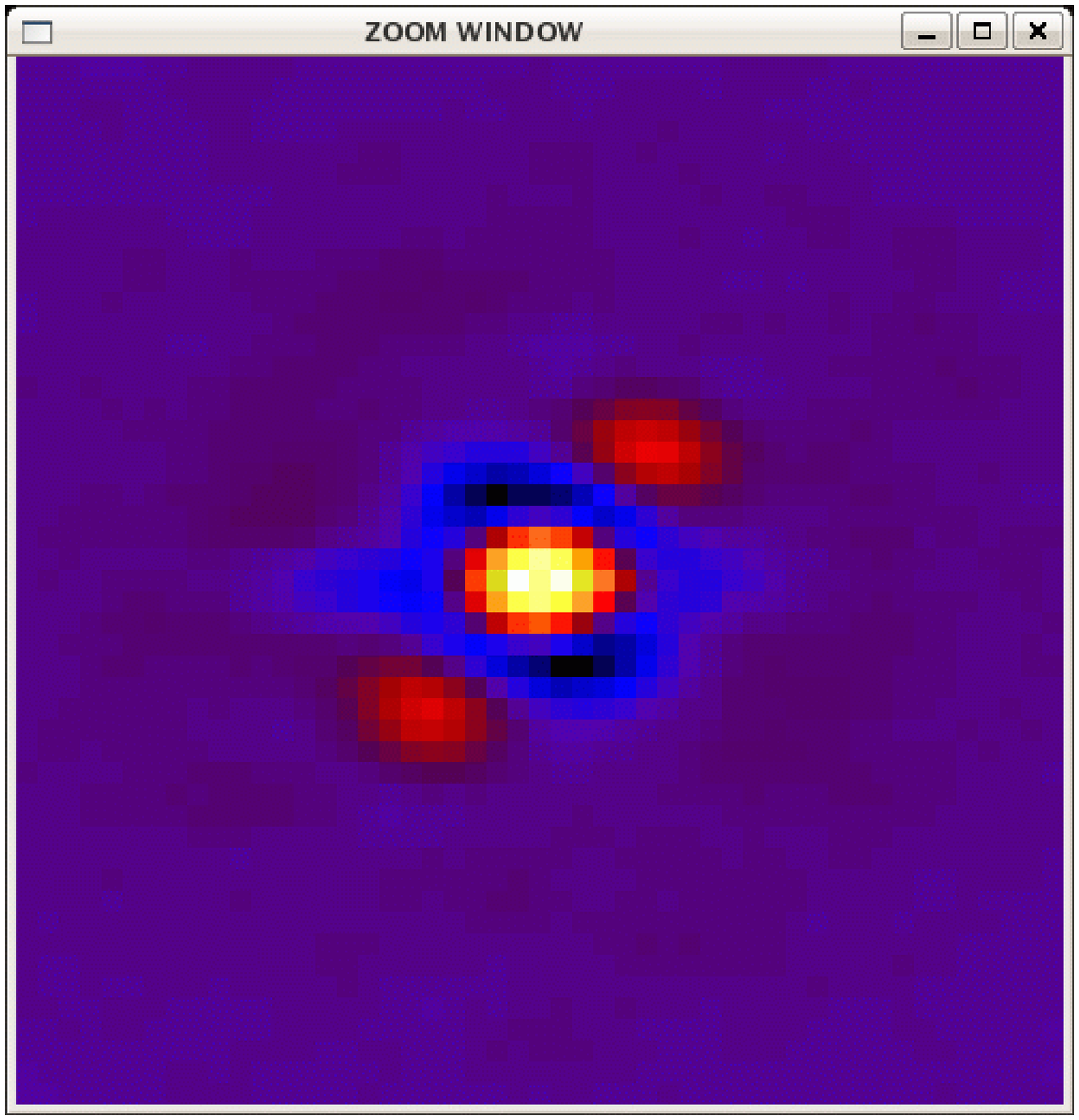}}
\caption{Panel a (left) is a directed vector autocorrelation of the subarray
around the brighter (Aa,Ab) pair of Finsen 332 while Panel b (right) is the 
fainter (Ba,Bb) pairing. These were generated from the same 1982.7650 data 
shown in Figure 3c. The images, at the same scale, provide a vivid 
representation of the pairs' similar morphologies.}
\label{fig4}
\end{figure}

\subsection{New Measures}

Additional observations made with the 4m telescopes of Kitt Peak National 
Observatory and Cerro Tololo Interamerican Observatory were obtained with 
the USNO speckle camera in 2001 and annually from 2005 to 2008. The system 
was also observed in 2006--07 with the Mt.\ Wilson 100in telescope and in 
2004 and 2008 with the Naval Observatory Flagstaff Station 61in telescope. 

Also re-reduced was an April 1996 observation with the CHARA ICCD of Aa,Ab. 
The observation has been initially inspected with no measure obtained 
(Ba,Bb was published in Hartkopf et al.\ 2000). Reanalysis of the archived 
videotape allowed this measure, at quite a close separation, to be made; 
results are included in Table 1.

\section{Discussion}

\subsection{Orbit Determination}

The larger errors associated with both micrometry and eyepiece 
interferometry, as well as the small $\Delta$m and the geometric 
peculiarities of the systems as illustrated in Figure 2, make them quite 
difficult to distinguish. However, observations by speckle interferometry 
are characterized by much lower errors (see Hartkopf et al.\ 2001). 
Therefore, only the measures obtained with 2m or larger telescopes are 
utilized in the orbit analyses. The measures not included in the orbit
determinations are indicated with notes in Table 1. The method of orbit 
calculation is the adaptive grid-search algorithm of Hartkopf et al.\ 
(1989), as modified by Mason et al.\ (1999). Briefly described, the 
Thiele-Innes elements (A,B, F and G) are calculated via an iterative 
three-dimensional grid-search of elements P, T$_{o}$, and e with the search 
parameter space decreasing as the elements converge. The remaining Campbell 
elements (a$''$, i, $\Omega$, and $\omega$) are then calculated directly. 
Observations are weighted using the scheme described in Hartkopf et al.\ 
(2001), which considers technique, observer expertise, the measured 
separation as a fraction of the telescope's Rayleigh limit, number of 
measures in a mean position, and any other notes the observer might have 
made with regard to quality. Table 4 gives the seven orbital elements along 
with their associated errors for both Aa,Ab and Ba,Bb. The degree of success
realized by the ensemble following the rubrics of Hartkopf et al.\ (2001), 
summarized by the grade, is also given here. Table 5 gives predicted 
positions ($\rho$ and $\theta$) at half-year intervals for the next five 
years. These orbits are illustrated in Figures 5 (Aa,Ab) and 6 (Ba,Bb).

\begin{figure}[!ht]
\epsfxsize 5.0in
\centerline{\epsffile{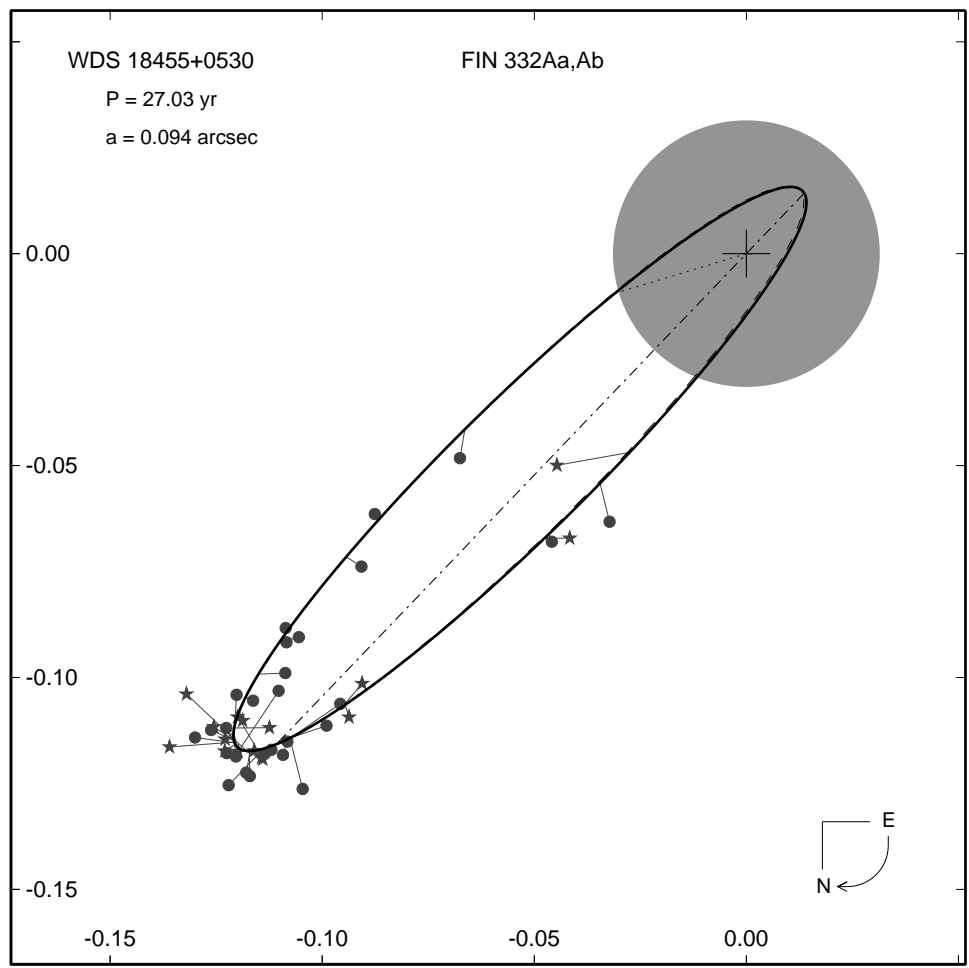}}
\caption{The relative orbit of FIN 332Aa,Ab, The different techniques are
represented by filled circles for published measures and filled stars for 
new or newly corrected measures. Only data used in this orbit determination
is plotted. The measures are connected to the predicted position by an O$-$C
line. The dashed line through the origin is the line of nodes. The light 
grey circle is the Rayleigh resolution limit ($\frac{1.22\lambda}{D}$) of a 
4m telescope. Unresolved measures from 4m class instruments are indicated by
a dotted line drawn from the origin. The scale is in arcseconds and the 
direction of motion is indicated in the lower right corner. The barely 
distinguishable dashed curve is the short period solution of Mason \& 
Hartkopf (2002).}
\label{fig5}
\end{figure}

\begin{figure}[!ht]
\epsfxsize 5.0in
\centerline{\epsffile{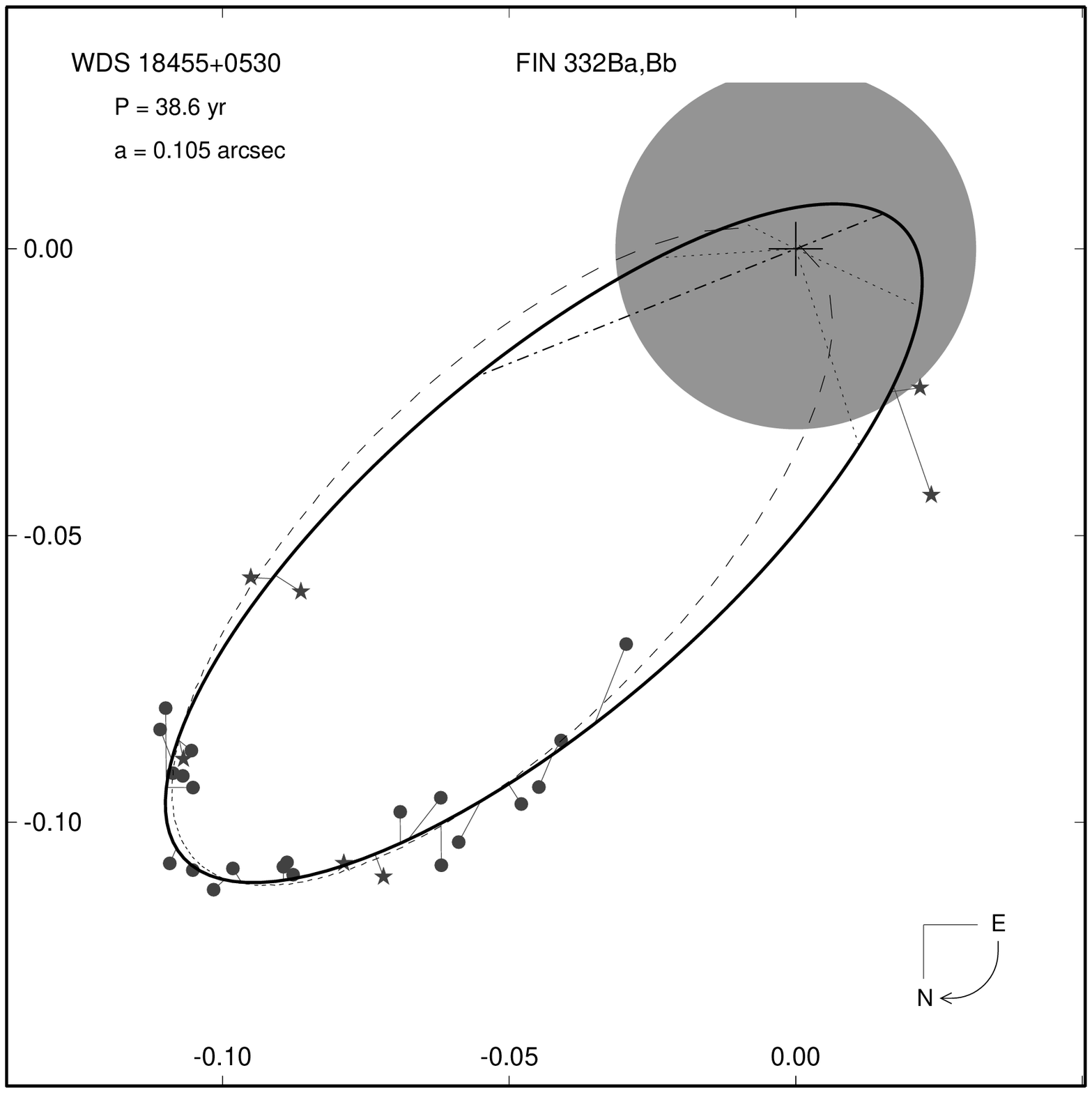}}
\caption{The relative orbit of FIN 332Ba,Bb. Symbols are the same as Figure 
5. Note the larger number of unresolved measures, and the greater divergence
from the Mason \& Hartkopf (2002) short period solution.}
\label{fig6}
\end{figure}

\subsubsection{Radial Velocity Measures}

One of the items of greatest interest to investigate was the initial mention
of radial velocity variability; however, this did not prove helpful in 
setting limits on orbital parameters. The spectral types of the components 
(A or a little later for each of them) makes the measurement of radial 
velocity variability quite difficult due to the broad nature of the spectral
features and the absence of many sharp metal lines. Plaskett et al.\ (1921) 
first noted variability and included STF2375 in their list of new 
spectroscopic binaries based on five observations obtained from June to 
October of 1920. Wilson (1953) added no new data in his catalog but gave it 
a quality rating of `acceptable.' Palmer et al.\ (1968) added eight new 
velocities, but changed the mean by only 1 km s$^{-1}$. Evans (1979) in his 
revision of Wilson's catalog later gave it a quality rating of `average.'

While most of the components in the multiple system are broad-lined A stars,
the Bb component may be an F star with sharper lines and it is possible 
that near periastron it may exhibit variable radial velocity features.

\subsubsection{Interpreting Unresolved Measures}

As seen in Tables 1 and 2 and illustrated in Figure 2, there are two recent 
times in the short-period orbits when the pairs were predicted to be 
unresolved: Aa,Ab from 1964.6 to 1969 and again from 1991.7 to 1996.1, Ba,Bb
from 1966.7 to 1972.1 and again from 2005.2 to 2010.7. The later two periods
of predicted non-resolution corresponded to multiple null detections for 
both pairs, as indicated in Tables 1 and 2. These non-resolutions, while not
utilized in determining these orbits, are completely consistent with the 
solutions.

\subsection{Mutual Inclination}

FIN 332 offers the rare possibility of determining the mutual inclination of
orbits in a quadruple system whose subsystems are at the same hierarchical 
level. A first glance shows that the individual orbital inclinations agree 
to with 1$\sigma$. However, the mutual inclination of their orbital planes 
is also dependent upon their nodal longitudes as given in the relation:

\[cos(\phi) = cos(i_{Aa,Ab})cos(i_{Ba,Bb}) + sin(i_{Aa,Ab})sin(i_{Ba,Bb})cos(\Omega_{Aa,Ab} - \Omega_{Ba,Bb}).\]

Inserting the values of (i,$\Omega$)$_{Aa,Ab}$ and (i,$\Omega$)$_{Ba,Bb}$ 
from Table 4 into this relation yields a mutual inclination of $\phi_{AB}$ =
25.2$\pm$12.2 degrees. This indicates that the two orbits are more coplanar 
than not; however, if we adopt the threshhold for coplanarity defined by 
Fekel (1981) of $\phi < 15^{\circ}$ then these two orbits are within 
1$\sigma$ of being coplanar. 

We have thus far assumed that the nodes specified by $\Omega_{Aa,Ab}$ and 
$\Omega_{Ba,Bb}$ are indeed the ascending nodes, but, regrettably, there is 
no spectroscopy to support that assumption. The two orbital inclinations 
reflect that both orbits are revolving in a retrograde sense, i.e. their 
position angles are decreasing with time. Interestingly, the wide and very 
long-period system is clearly moving in a direct sense with position angles 
increasing with time.

Because of the degeneracy of the Omegas, a second possible value for the 
mutual inclination of 49.3$\pm$19.6 degrees arises. While the eye is 
naturally attracted to the case of identical nodal quadrants, without radial
velocity verification, which has already been shown to be a very challenging
task, there remains the possibility that nature lacks the aesthetic of the 
eye. At this point, all we can state is that while the two orbital planes 
may be nearly coplanar, they are most certainly not nearly perpendicular.

\subsection{Mass Sums}

While both Aa and Ba are listed as spectral type A1V in the Multiple Star 
Catalog (Tokovinin 1997), the spectral types of the secondaries are not 
known. Given the small magnitude differences (discounting the Hipparcos 
$\Delta$m) it is conceivable that we have four A dwarf stars with expected 
mass sums of each pair between 5 and 6 $\cal{M}_{\odot}$. Unfortunately, 
given the large errors in the parallax and orbital elements these are of 
little help. The Aa,Ab solution gives a mass sum of 
$12\pm$16$\cal{M}_{\odot}$ while that of Ba,Bb is 
$7.6\pm9.0\cal{M}_{\odot}$. While their orbital elements can undoubtedly be 
improved, especially if they are resolved during periastron, the largest 
improvement may come from a more precise determination of their parallax.

\subsection{Stitching `Dum' and `Dee' on their collars}

For the first several decades since their discovery, the peculiar geometries
of these systems made them nearly indistinguishable. If we compare their 
predicted position and subjectively qualify them as $``$similar'' when their
positions are approximately the same: d$\theta$~$<$10$^{\circ}$ and 
d$\rho$~$<$~0\farcs05 or both $<$~0\farcs05, i.e., unresolved, they would be
qualified as $``$similar" for 33\% of the next thousand years. While their 
appearances have diverged somewhat in recent years, by the middle of this 
century both pairs will again go through periastron within a few years of 
each other and FIN 332 Aa,Ab and Ba,Bb will again exemplify their Carrollian
sobriquets. 

\acknowledgments
 
We would like to acknowledge William S. Finsen and William van den Bos for
their pioneering work in double star interferometry in the southern 
hemisphere. Also we would like to thank the USNO on-site archives for access
to the voluminous correspondence between W.S. Finsen and C.E. Worley, head 
of the USNO visual double star program and Washington Double Star catalog 
projects for more than thirty years. The USNO speckle interferometry program
has been supported by the National Aeronautics and Space Administration 
under reimburseable \#NNH06AD70I, issued through the Terrestrial Planet 
Finder Foundation Science program. This research has made use of the SIMBAD 
database, operated at CDS, Strasbourg, France. Thanks are also extended to 
Ken Johnston and the U.\ S.\ Naval Observatory for their continued support 
of the Double Star Program.

\newpage

%\documentstyle[aj_pt4]{article}
%\begin{document}
%\pagestyle{empty} 
\begin{deluxetable}{ccccccclc}

\tabletypesize{\footnotesize}

\newdimen\digitwidth
\setbox0=\hbox{\footnotesize 2}
\digitwidth=\wd0
\catcode`#=\active
\def#{\kern\digitwidth}

\newdimen\ltwidth
\setbox0=\hbox{\footnotesize $<$}
\ltwidth=\wd0
\catcode`@=\active
\def@{\kern\ltwidth}

\newdimen\colonwidth
\setbox0=\hbox{\footnotesize :}
\colonwidth=\wd0
\catcode`!=\active
\def!{\kern\colonwidth}

\tablenum{1}
\tablewidth{0pt}
\tablecaption{Measurements of FIN 332Aa,Ab}
\tablehead{
\colhead{Epoch} &
\colhead{$\theta$} & 
\colhead{$\rho$} &
\colhead{n} &
\colhead{O$-$C} &
\colhead{O$-$C} &
\colhead{Method} &
\colhead{Reference} &
\colhead{Notes} \nl
\colhead{~} & 
\colhead{($\circ$)} & 
\colhead{($''$)} &
\colhead{~} &
\colhead{($\circ$)} & 
\colhead{($''$)} & 
\colhead{~} &
\colhead{~} &
\colhead{~} 
}
\startdata
1953.73## &  316.5     &    0.153          & 5 &    #3.0   & $-$0.014@ & E        & Finsen 1953                & 1     \nl
1953.74## &  315.7     &    0.15#          & 1 &    #2.2   & $-$0.017@ & M        & van den Bos 1956           & 1     \nl
1954.68## &  302.7     &    0.158          & 4 & $-$10.2@  & $-$0.007@ & E        & Finsen 1956                & 1     \nl
1955.72## &  309.8     &    0.144          & 3 & #$-$2.3@  & $-$0.017@ & E        & Finsen 1956                & 1     \nl
1957.39## &  311.9     &    0.15#          & 4 &    #1.2   & $-$0.002@ & M        & van den Bos 1958b          & 1     \nl
1957.76## &  314.7     &    0.144          & 1 &    #4.3   & $-$0.005@ & E        & Finsen 1959                & 1     \nl
1957.89## &  314.6     &    0.13#          & 4 &    #4.3   & $-$0.018@ & M        & van Biesbroeck 1960        & 1     \nl
1958.54## &  311.6     &    0.15#          & 3 &    #1.9   &    0.008  & M        & van den Bos 1960           & 1     \nl
1959.72## &  302.9     &    0.131          & 3 & #$-$5.6@  &    0.000  & E        & Finsen 1960                & 1     \nl
1960.564# &  318.2     &    0.14#          & 6 &    10.8   &    0.019  & M        & van Biesbroeck 1965        & 1     \nl
1960.72## &  298.9     &    0.137          & 1 & #$-$8.3@  &    0.018  & E        & Finsen 1961                & 1     \nl
1961.57## &  312.6     &    0.11#          & 5 &    #6.7   &    0.003  & M        & van den Bos 1962           & 1     \nl
1961.73## &  297.8     &    0.112          & 3 & #$-$7.8@  &    0.008  & E        & Finsen 1962                & 1     \nl
1962.51## &  314.7     &    0.10#          & 4 &    10.7   &    0.008  & M        & van den Bos 1963a          & 1     \nl   
1962.72## &  309.2     &    0.114          & 5 &    #5.7   &    0.026  & E        & Finsen 1963                & 1     \nl
1963.38## & \multicolumn{2}{c}{unresolved} & 1 &  (301.5)# &   (0.076) & M        & van den Bos 1963b          & 1,2   \nl  
1963.728# &  313.0     &    0.106          & 4 &    12.8   &    0.037  & E        & Finsen 1964a               & 1     \nl
1964.726# & \multicolumn{2}{c}{unresolved} & 1 &  (294.1)# &   (0.046) & E        & Finsen 1965                & 1,2   \nl
1966.758# & \multicolumn{2}{c}{unresolved} & 1 &  (157.0)# &   (0.016) & E        & Finsen 1967                & 1,2   \nl
1968.791# & \multicolumn{2}{c}{unresolved} & 1 &  (334.2)# &   (0.040) & E        & Finsen 1969                & 1,2   \nl
1971.531# &  307.0     &    0.15#          & 1 & $-$15.2@  &    0.046  & M        & Walker 1972                & 1     \nl
1975.48## &  316.9     &    0.12#          & 3 & #$-$0.6@  & $-$0.031@ & M        & Heintz 1978                & 1     \nl
1976.2992 &  318.1     &    0.143          & 1 &    #1.2   & $-$0.014@ & Sp       & McAlister 1978             &       \nl
1976.3702 &  318.5     &    0.149          & 1 &    #1.7   & $-$0.008@ & Sp       & McAlister \& Hendry 1982a  &       \nl
1976.3728 &  320.5     &    0.164          & 1 &    #3.7   &    0.007  & Sp       & McAlister \& Hendry 1982a  &       \nl
1976.4549 &  317.4     &    0.161          & 1 &    #0.6   &    0.004  & Sp       & McAlister 1978             &       \nl
1977.3340 &  316.9     &    0.158          & 1 &    #0.8   & $-$0.004@ & Sp       & McAlister \& Hendry 1982a  &       \nl
1977.4815 &  316.4     &    0.162          & 1 &    #0.4   & $-$0.000@ & Sp       & McAlister 1979             &       \nl
1977.4871 &  316.2     &    0.164          & 1 &    #0.2   &    0.002  & Sp       & McAlister 1979             &       \nl
1977.521# &  312.9     &    0.18#          & 3 & #$-$2.9@  & $-$0.017@ & M        & Walker 1985                & 1     \nl
1977.6400 &  315.9     &    0.175          & 1 &    #0.0   &    0.012  & Sp       & McAlister \& Hendry 1982a  &       \nl
1978.5410 &  316.2     &    0.170          & 1 &    #1.0   &    0.004  & Sp       & McAlister \& Fekel 1980    &       \nl
1978.6147 &  316.6     &    0.170          & 1 &    #1.4   &    0.004  & Sp       & McAlister \& Fekel 1980    &       \nl
1979.3601 &  314.0     &    0.170          & 1 & #$-$0.6@  &    0.003  & Sp       & McAlister \& Hendry 1982b  &       \nl
1979.5321 &  313.2     &    0.151          & 1 & #$-$1.3@  & $-$0.016@ & Sp       & McAlister \& Hendry 1982b  &       \nl
1979.7725 &  312.5     &    0.166          & 1 & #$-$1.8@  & $-$0.001@ & Sp       & McAlister \& Hendry 1982b  &       \nl
1980.4769 &  311.4     &    0.173          & 1 & #$-$2.4@  &    0.006  & Sp       & McAlister et al.\ 1983     &       \nl
1980.4794 &  314.7     &    0.169          & 1 &    #0.9   &    0.002  & Sp       & McAlister \& Hartkopf 1984 &       \nl
1980.7173 &  311.0     &    0.159          & 1 & #$-$2.7@  & $-$0.008@ & Sp       & McAlister et al.\ 1983     &       \nl
1980.7199 &  311.8     &    0.169          & 1 & #$-$1.9@  &    0.002  & Sp       & McAlister et al.\ 1983     &       \nl
1981.356# &  313.2     &    0.186          & 1 &    #0.1   &    0.020  & P        & Tokovinin 1982             & 1     \nl
1982.5029 &  316.4     &    0.165          & 1 &    #4.0   &    0.003  & Sc       & This paper                 & 3     \nl
1982.5248 &  315.2     &    0.160          & 1 &    #3.0   & $-$0.002@ & Sc       & Fu et al.\ 1997            & 1     \nl
1982.7650 &  313.0     &    0.162          & 2 &    #0.8   &    0.001  & Sc       & This paper                 & 3     \nl
1983.4203 &  312.3     &    0.157          & 1 &    #0.7   &    0.001  & Sc       & McAlister et al.\ 1987a    & 4     \nl
1984.3760 &  312.4     &    0.147          & 1 &    #1.5   & $-$0.005@ & Sc       & Hartkopf et al.\ 2000      & 5     \nl
1984.783# &  335.9     &    0.127          & 1 &    25.5   & $-$0.022@ & P        & Tokovinin \& Ismailov 1988 & 1     \nl
1985.4816 &  310.7     &    0.139          & 1 &    #0.8   & $-$0.004@ & Sc       & McAlister et al.\ 1987a    & 4     \nl
1985.5231 &  310.3     &    0.142          & 1 &    #0.5   & $-$0.001@ & Sc       & McAlister et al.\ 1987b    & 4     \nl
1985.7440 &  318.3     &    0.137          & 1 &    #8.8   & $-$0.004@ & P        & Tokovinin \& Ismailov 1988 & 1     \nl
1985.8424 &  309.2     &    0.140          & 1 & #$-$0.3@  &    0.000  & Sc       & McAlister et al.\ 1987a    & 4     \nl
1987.7618 &  309.2     &    0.117          & 1 &    #2.0   & $-$0.001@ & Sc       & McAlister et al.\ 1989     & 4     \nl
1988.6655 &  305.1     &    0.107          & 1 & #$-$0.7@  &    0.001  & Sc       & McAlister et al.\ 1990     &       \nl
1990.2734 &  305.6     &    0.083          & 1 &    #3.6   &    0.005  & Sc       & Hartkopf et al.\ 1992      &       \nl
1991.2500 & \multicolumn{2}{c}{unresolved} &   &  (297.8)# &   (0.058) & H        & ESA 1997                   & 2,6   \nl
1992.3105 &            & $<$0.038@         & 1 &  (286.8)# &   (0.032) & Sc       & This Paper                 & 2,7   \nl
1996.3214 &  318.2     &    0.067          & 1 & $-$11.4@  &    0.013  & Sc       & This Paper                 & 8     \nl
1996.6930 &  333.0     &    0.071          & 1 &    #5.5   &    0.007  & Sc       & Hartkopf et al.\ 2000      &       \nl
1997.3945 &  326.0     &    0.082          & 1 &    #1.1   &    0.001  & S        & Balega et al.\ 1999        &       \nl
1997.4630 &  328.2     &    0.079          & 1 &    #3.5   & $-$0.003@ & Sc       & This Paper                 & 9     \nl
2001.4988 &  319.4     &    0.144          & 1 &    #1.0   &    0.002  & S$\star$ & This Paper                 & 10    \nl
2001.5697 &  318.2     &    0.136          & 1 & #$-$0.1@  & $-$0.007@ & S$\star$ & This Paper                 & 11    \nl
2005.8652 &  308.2     &    0.168          & 1 & #$-$6.7@  &    0.002  & S$\star$ & This Paper                 & 10    \nl
2006.2001 &  315.3     &    0.165          & 1 &    #0.7   & $-$0.002@ & S$\star$ & This Paper                 & 11    \nl
2006.5640 &  316.1     &    0.165          & 1 &    #1.8   & $-$0.002@ & S$\star$ & This Paper                 & 9     \nl
2007.3174 &  313.6     &    0.170          & 1 & #$-$0.2@  &    0.003  & S$\star$ & This Paper                 & 9     \nl
2007.5879 &  310.5     &    0.179          & 2 & #$-$3.1@  &    0.012  & S$\star$ & This Paper                 & 10    \nl
2007.8010 &  311.6     &    0.168          & 1 & #$-$1.9@  &    0.001  & S$\star$ & This Paper                 & 9     \nl
2008.4529 &  313.0     &    0.168          & 4 & #$-$0.1@  &    0.003  & S$\star$ & This Paper                 & 10    \nl
2008.5371 &  314.4     &    0.168          & 2 &    #1.5   &    0.003  & S        & Tokovinin et al.\ 2010     & 12    \nl
2008.7721 &  314.8     &    0.159          & 1 &    #2.0   & $-$0.006@ & S        & Tokovinin et al.\ 2010     & 13    \nl
2008.8712 &  316.9     &    0.192          & 2 &    #4.2   &    0.028  & S$\star$ & This Paper                 & 1, 14 \nl
2009.2607 &  312.3     &    0.162          & 2 & #$-$0.1@  & $-$0.001@ & S        & Tokovinin et al.\ 2010     & 12    \nl
\enddata
\tablenotetext{~}{Methods: E = eyepiece interferometer, H = Hipparcos observation, M = micrometer, P = phase grating
                  interferometer, S = speckle interferometer, Sp = photographic speckle camera of McAlister (1977), Sc = 
                  ICCD speckle camera of McAlister et al.\ (1987a), S$\star$ = USNO speckle camera of Mason et al.\ (2009).}
\tablenotetext{~}{#1 : Measure not used in new orbit solution.}                         
\tablenotetext{~}{#2 : Here Columns 4 \& 5 give the predicted position of the secondary relative to the primary.}
\tablenotetext{~}{#3 : Measure obtained by re-reduction of CCD subarray. See \S 4.4.}
\tablenotetext{~}{#4 : The original calibration was corrected in McAlister et al.\ (1989) and this corrected measure first
                       published in McAlister \& Hartkopf (1988).}
\tablenotetext{~}{#5 : Re-reduction of data yielded improved SNR and allowed this measure to be made.}
\tablenotetext{~}{#6 : No measure of this subsystem was published in the Hipparcos Catalogue.}
\tablenotetext{~}{#7 : The other pair, Ba,Bb (see Table 2) was measured at this time, so this is judged to be a reliable
                       null detection.}
\tablenotetext{~}{#8 : Measure inadvertently left out of Hartkopf et al.\ (2000).}
\tablenotetext{~}{#9 : Observation made on Mt.\ Wilson 100$''$.}
\tablenotetext{~}{10 : Observation made on KPNO 4m.}
\tablenotetext{~}{11 : Observation made on CTIO 4m.}
\tablenotetext{~}{12 : $\Delta$m is 0.9$\pm$0.4 in Str\"{o}mgren y.}
\tablenotetext{~}{13 : $\Delta$m is 1.3 in H$\alpha$.}
\tablenotetext{~}{14 : Observation made on NOFS 61$''$.}
\end{deluxetable}
%\end{document}

%\documentstyle[aj_pt4]{article}
%\begin{document}
%\pagestyle{empty} 

\begin{deluxetable}{ccccccclc}
\tabletypesize{\footnotesize}

\newdimen\digitwidth
\setbox0=\hbox{\footnotesize 2}
\digitwidth=\wd0
\catcode`#=\active
\def#{\kern\digitwidth}

\newdimen\ltwidth
\setbox0=\hbox{\footnotesize $<$}
\ltwidth=\wd0
\catcode`@=\active
\def@{\kern\ltwidth}

\newdimen\colonwidth
\setbox0=\hbox{\footnotesize :}
\colonwidth=\wd0
\catcode`!=\active
\def!{\kern\colonwidth}

\tablenum{2}
\tablewidth{0pt}
\tablecaption{Measurements of FIN 332Ba,Bb}
\tablehead{
\colhead{Epoch} &
\colhead{$\theta$} & 
\colhead{$\rho$} &
\colhead{n} &
\colhead{O$-$C} &
\colhead{O$-$C} &
\colhead{Method} &
\colhead{Reference} &
\colhead{Notes} \nl
\colhead{~} & 
\colhead{($\circ$)} & 
\colhead{($''$)} &
\colhead{~} &
\colhead{($\circ$)} & 
\colhead{($''$)} & 
\colhead{~} &
\colhead{~} &
\colhead{~} 
}
\startdata
1953.73## & 315.3 &    0.148               & 5 &    #1.3   & $-$0.002@ & E        & Finsen 1953                & 1       \nl
1953.74## & 315.1 &    0.14#               & 1 &    #1.1   & $-$0.010@ & M        & van den Bos 1956           & 1       \nl
1954.68## & 317.6 &    0.144               & 4 &    #4.6   & $-$0.005@ & E        & Finsen 1956                & 1       \nl
1955.72## & 314.9 &    0.141               & 3 &    #3.0   & $-$0.007@ & E        & Finsen 1956                & 1       \nl
1957.39## & 310.0 &    0.16#               & 4 & #$-$0.1@  &    0.017  & M        & van den Bos 1958           & 1       \nl
1957.73## & 300.# &    0.15#               & 2 & #$-$9.7@  &    0.008  & M        & Muller 1958                & 1       \nl
1957.73## & 306.# &    0.14#               & 1 & #$-$3.7@  & $-$0.002@ & M        & Muller 1958                & 1       \nl
1957.76## & 308.3 &    0.147               & 1 & #$-$1.4@  &    0.005  & E        & Finsen 1959                & 1       \nl
1957.89## & 313.8 &    0.12#               & 4 &    #4.3   & $-$0.021@ & M        & van Biesbroeck 1960        & 1       \nl
1958.54## & 312.3 &    0.15#               & 3 &    #3.6   &    0.011  & M        & van den Bos 1960           & 1       \nl
1959.72## & 312.3 &    0.124               & 3 &    #5.1   & $-$0.009@ & E        & Fin1960b                   & 1       \nl
1960.564# & 312.4 &    0.13#               & 6 &    #5.8   &    0.003  & M        & van Biesbroeck 1965        & 1       \nl
1960.72## & 310.9 &    0.139               & 1 &    #5.1   &    0.013  & E        & Finsen 1961                & 1       \nl
1961.57## & 311.2 &    0.13#               & 5 &    #6.7   &    0.010  & M        & van den Bos 1962           & 1       \nl
1961.73## & 311.0 &    0.126               & 3 &    #6.7   &    0.007  & E        & Finsen 1962                & 1       \nl
1962.51## & 312.0 &    0.11#               & 4 &    #9.1   & $-$0.002@ & M        & van den Bos 1963a          & 1       \nl
1962.72## & 320.8 &    0.123               & 5 &    18.3   &    0.014  & E        & Finsen 1963                & 1       \nl
1963.38## & \multicolumn{2}{c}{unresolved} & 1 &  (301.1)# &   (0.102) & M        & van den Bos 1963b          & 1,2     \nl
1963.728# & 323.6 &    0.113               & 4 &    23.3   &    0.015  & E        & Finsen 1964a               & 1       \nl
1964.726# & \multicolumn{2}{c}{unresolved} & 1 &  (297.5)# &   (0.084) & E        & Finsen 1965                & 1,2     \nl
1966.436# & 276.7 &    0.26#               & 1 & $-$11.7@  &    0.210  & M        & Walker 1969                & 1       \nl
1966.758# & \multicolumn{2}{c}{unresolved} & 1 &  (284.8)# &   (0.041) & E        & Finsen 1967                & 1,2     \nl
1968.791# & \multicolumn{2}{c}{unresolved} & 1 &  (#72.1)# &   (0.023) & E        & Finsen 1969                & 1,2     \nl
1971.504# & #90.0 &    0.15#               & 1 &    84.9   &    0.105  & M        & Walker 1972                & 1       \nl
1976.4549 & 336.9 &    0.075               & 1 & #$-$0.3@  & $-$0.015@ & Sp       & McAlister 1978             &         \nl
1977.4815 & 334.6 &    0.095               & 1 &    #0.2   & $-$0.003@ & Sp       & McAlister 1979             &         \nl
1977.4870 & 334.6 &    0.104               & 1 &    #0.2   &    0.006  & Sp       & McAlister 1979             &         \nl
1977.521# & 317.0 &    0.12#               & 3 & $-$17.2@  &    0.022  & M        & Walker 1985                & 1       \nl
1978.6147 & 333.8 &    0.108               & 1 &    #1.9   &    0.002  & Sp       & McAlister \& Fekel 1980    &         \nl
1979.3601 & 330.5 &    0.119               & 1 &    #0.1   &    0.008  & Sp       & McAlister \& Hendry 1982b  &         \nl
1980.4769 & 330.2 &    0.124               & 1 &    #1.8   &    0.006  & Sp       & McAlister et al.\ 1983     &         \nl
1981.356# & 321.5 &    0.111               & 1 & #$-$5.4@  & $-$0.012@ & P        & Tokovinin 1982             & 1       \nl
1981.4681 & 327.2 &    0.114               & 1 &    #0.4   & $-$0.009@ & Sp       & McAlister et al.\ 1984     &         \nl
1981.6975 & 325.0 &    0.120               & 1 & #$-$1.5@  & $-$0.005@ & Sp       & McAlister et al.\ 1984     &         \nl
1982.5029 & 326.8 &    0.131               & 1 &    #1.5   &    0.002  & Sc       & This Paper                 & 3       \nl
1982.7650 & 323.7 &    0.133               & 1 & #$-$1.2@  &    0.003  & Sc       & This Paper                 & 3       \nl
1984.783# & 329.4 &    0.103               & 1 &    #7.2   & $-$0.035@ & P        & Tokovinin \& Ismailov 1988 & 1,4     \nl
1985.4816 & 321.3 &    0.140               & 1 & #$-$0.1@  & $-$0.001@ & Sc       & McAlister et al.\ 1987a    & 5       \nl
1985.5231 & 320.4 &    0.139               & 1 & #$-$1.0@  & $-$0.002@ & Sc       & McAlister et al.\ 1987b    & 5       \nl
1985.7440 & 308.3 &    0.100               & 1 & $-$12.8@  & $-$0.042@ & P        & Tokovinin \& Ismailov 1988 & 1,4     \nl
1985.8424 & 320.4 &    0.140               & 1 & #$-$0.6@  & $-$0.002@ & Sc       & McAlister et al.\ 1987a    & 5       \nl
1987.7618 & 317.8 &    0.146               & 1 & #$-$1.1@  & $-$0.001@ & Sc       & McAlister et al.\ 1989     &         \nl
1988.6655 & 317.8 &    0.151               & 1 & #$-$0.1@  &    0.003  & Sc       & McAlister et al.\ 1990     &         \nl
1990.2734 & 315.9 &    0.151               & 1 & #$-$0.3@  &    0.001  & Sc       & Hartkopf et al.\ 1992      &         \nl
1991.2500 & 308.# &    0.16#               & 0 & #$-$7.1@  & $-$0.009@ & H        & ESA 1997                   & 1,6     \nl
1992.3105 & 314.5 &    0.153               & 1 &    #0.4   &    0.003  & Sc       & Hartkopf et al.\ 1994      &         \nl
1995.6008 & 306.1 &    0.136               & 1 & #$-$4.5@  & $-$0.009@ & Sc       & Hartkopf et al.\ 1997      & 7       \nl
1995.6061 & 311.8 &    0.141               & 1 &    #1.2   & $-$0.004@ & Sc       & Hartkopf et al.\ 2000      &         \nl
1996.3215 & 310.7 &    0.141               & 1 &    #1.0   & $-$0.001@ & Sc       & Hartkopf et al.\ 2000      &         \nl
1996.3270 & 310.1 &    0.142               & 1 &    #0.4   & $-$0.000@ & Sc       & Hartkopf et al.\ 2000      &         \nl
1996.7012 & 307.1 &    0.139               & 1 & #$-$2.2@  & $-$0.002@ & Sc       & Hartkopf et al.\ 2000      &         \nl
1997.3945 & 309.7 &    0.137               & 1 &    #1.2   & $-$0.001@ & S        & Balega et al.\ 1999        &         \nl
1997.4630 & 309.8 &    0.139               & 1 &    #1.4   &    0.002  & Sc       & This Paper                 & 8       \nl
2001.4988 & 301.1 &    0.111               & 3 & #$-$1.1@  &    0.003  & S$\star$ & This Paper                 & 9       \nl
2001.5697 & 304.7 &    0.105               & 1 &    #2.6   & $-$0.002@ & S$\star$ & This Paper                 & 10      \nl
2005.8652 &       & $<$0.038@              & 1 &  (273.7)# &   (0.024) & S$\star$ & This Paper                 & 2,9,11  \nl
2006.2001 &       & $<$0.038@              & 1 &  (243.3)# &   (0.011) & S$\star$ & This Paper                 & 2,10,11 \nl
2006.5640 &       & $<$0.060@              & 1 &  (127.2)# &   (0.013) & S$\star$ & This Paper                 & 2,11,12 \nl
2007.3174 &       & $<$0.060@              & 1 &  (#76.7)# &   (0.023) & S$\star$ & This Paper                 & 2,11,12 \nl
2007.5879 &       & $<$0.038@              & 1 &  (#65.0)# &   (0.024) & S$\star$ & This Paper                 & 2,9,11  \nl
2007.8010 &       & $<$0.060@              & 1 &  (#56.7)# &   (0.025) & S$\star$ & This Paper                 & 2,11,12 \nl
2008.4615 & #28.8 &    0.049               & 2 & #$-$6.7@  &    0.020  & S$\star$ & This Paper                 & 9       \nl
2008.5371 & #41.8 &    0.033               & 1 &    #8.4   &    0.003  & S        & Tokovinin et al.\ 2010     & 13      \nl
2008.8658 &       & $<$0.098@              & 1 &  (#25.6)# &   (0.033) & S$\star$ & This Paper                 & 2,14    \nl
2009.2607 &       & $<$0.050@              & 1 &  (#17.8)# &   (0.036) & S        & Tokovinin et al.\ 2010     & 2,15    \nl
\enddata
\tablenotetext{~}{Methods: E = eyepiece interferometer, H = Hipparcos observation, M = micrometer, P = phase grating
                  interferometer, S = speckle interferometer, Sp = photographic speckle camera of McAlister (1977), Sc =
                  ICCD speckle camera of McAlister et al.\ (1987a), S$\star$ = USNO speckle camera of Mason et al.\ (2009).}
\tablenotetext{~}{#1 : Measure not used in new orbit solution.}                         
\tablenotetext{~}{#2 : Here Columns 4 \& 5 give the predicted position of the secondary relative to the primary.}
\tablenotetext{~}{#3 : Measure obtained by re-reduction of CCD subarray. See \S 4.4.}
\tablenotetext{~}{#4 : Published position angle was 59\fdg4, and 38\fdg3 and given zero weight in orbit determination. See
                       \S 4.1.}
\tablenotetext{~}{#5 : The original calibration was corrected in McAlister et al.\ (1989) 
                       and this corrected measure first published in McAlister \& Hartkopf (1988).}
\tablenotetext{~}{#6 : The H$_{p}$ magnitude difference is 0.76$\pm$0.15.}
\tablenotetext{~}{#7 : Assigned in error to Aa,Ab in Hartkopf et al.\ (1997).}
\tablenotetext{~}{#8 : Observation made on Mt.\ Wilson 100$''$.}
\tablenotetext{~}{#9 : Observation made on KPNO 4m.}
\tablenotetext{~}{10 : Observation made on CTIO 4m.}
\tablenotetext{~}{11 : The other pair, Aa,Ab (see Table 1) was measured at this time, so this is judged to be a reliable
                       null detection.}
\tablenotetext{~}{12 : Observation made on Mt.\ Wilson 100$''$. Not plotted in Figure 6.}
\tablenotetext{~}{13 : $\Delta$m is 0.5 in Str\"{o}mgren y.}
\tablenotetext{~}{14 : Observation made on NOFS 61$''$. Not plotted in Figure 6.}
\tablenotetext{~}{15 : Observation obtained on the SOAR 4.2m telescope. While Ba,Bb was previously resolved when it was
                       closer according to A. Tokovinin: ``Bab could be partially resolvable, but in the AD [Atmosheric
                       Dispersion] direction. Fits do not converge, so it remains unresolved. The AD was 3.2 pixels, so if
                       the pair was under 50mas or so, the negative result could be explained."}
\end{deluxetable}
%\end{document}

%\documentstyle[aj_pt4]{article}
%\begin{document}
%\pagestyle{empty} 
\footnotesize

\newdimen\digitwidth
\setbox0=\hbox{2}
\digitwidth=\wd0
\catcode`#=\active
\def#{\kern\digitwidth}

\newdimen\ltwidth
\setbox0=\hbox{$<$}
\ltwidth=\wd0
\catcode`@=\active
\def@{\kern\ltwidth}

\newdimen\colonwidth
\setbox0=\hbox{:}
\colonwidth=\wd0
\catcode`!=\active
\def!{\kern\colonwidth}

\begin{deluxetable}{cccccc}
\tablenum{3}
\tablewidth{0pt}
\tablecaption{Measurements of STF2375AB}
\tablehead{
\colhead{Epoch} &
\colhead{$\theta$} & 
\colhead{$\rho$} &
\colhead{n} &
\colhead{Method} &
\colhead{Notes} \nl
\colhead{~} & 
\colhead{($\circ$)} & 
\colhead{($''$)} &
\colhead{~} 
}
\startdata
1997.4657 & 119.6 & 2.590 & 1 & Sc       & 1 \nl
2004.2019 & 122.9 & 2.496 & 1 & S$\star$ & 2 \nl
2006.1974 & 120.1 & 2.549 & 1 & S$\star$ & 3 \nl
2006.5640 & 119.7 & 2.512 & 1 & S$\star$ & 1 \nl
2007.3174 & 119.6 & 2.484 & 1 & S$\star$ & 1 \nl
2007.5879 & 118.2 & 2.537 & 1 & S$\star$ & 4 \nl
2007.8010 & 118.2 & 2.537 & 1 & S$\star$ & 1 \nl
2008.4569 & 119.5 & 2.504 & 3 & S$\star$ & 4 \nl
2008.8549 & 119.1 & 2.618 & 3 & S$\star$ & 2 \nl
2008.8712 & 119.1 & 2.569 & 3 & S$\star$ & 2 \nl
\enddata
\tablenotetext{~}{Methods: Sc = ICCD speckle camera of McAlister et al.\ (1987a), 
                  S$\star$ = USNO speckle camera of Mason et al.\ (2009)}
\tablenotetext{~}{#4 : Observation made on KPNO 4m.}
\tablenotetext{~}{#3 : Observation made on CTIO 4m.}
\tablenotetext{~}{#1 : Observation made on Mt.\ Wilson 100$''$.}
\tablenotetext{~}{#2 : Observation made on NOFS 61$''$.}
\end{deluxetable}
%\end{document}

%\documentstyle[aj_pt4]{article}
%\begin{document}
%\setlength{\textwidth}{7.0in}
%\setlength{\textheight}{9.5in}
\pagestyle{empty} 
%\footnotesize
%\small
\newdimen\digitwidth
\setbox0=\hbox{2}
\digitwidth=\wd0
\catcode`#=\active
\def#{\kern\digitwidth}

\newdimen\ltwidth
\setbox0=\hbox{$<$}
\ltwidth=\wd0
\catcode`@=\active
\def@{\kern\ltwidth}

\newdimen\colonwidth
\setbox0=\hbox{:}
\colonwidth=\wd0
\catcode`!=\active
\def!{\kern\colonwidth}

\begin{deluxetable}{lcc}
\tablenum{4}
\tablewidth{0pt}
\tablecaption{Orbital Elements of FIN 332Aa,Ab \& Ba,Bb}
\tablehead{
\colhead{Element} &
\colhead{FIN 332Aa,Ab} & 
\colhead{FIN 332Ba,Bb} }
\startdata
Period; P (yrs)                                & ##27.03#$\pm$#0.67# & ##38.6##$\pm$#1.2## \nl
Semi-major axis; a$''$                         & ###0.094$\pm$#0.019 & ###0.105$\pm$#0.015 \nl
Inclination; i ($^{\circ}$)                    & #106.###$\pm$20.### & #117.2##$\pm$#9.5## \nl
Longitude of Node; $\Omega$ ($^{\circ}$)       & #136.2##$\pm$#4.2## & #111.8##$\pm$#5.7## \nl
Epoch of Periastron; T$_{o}$ (yrs)             & 1994.20#$\pm$#0.98# & 1967.9##$\pm$#1.9## \nl
Eccentricity; e                                & ###0.79#$\pm$#0.34# & ###0.867$\pm$#0.034 \nl
Longitude of Periastron; $\omega$ ($^{\circ}$) & ##10.###$\pm$16.### & #311.2##$\pm$#8.3## \nl
\vspace{-5pt} \nl
Grade                                          & 3                   & 3                   \nl
\enddata
\end{deluxetable}
%\end{document}

%\documentstyle[aj_pt4]{article}
%\begin{document}
%\setlength{\textwidth}{7.0in}
%\setlength{\textheight}{9.5in}
\pagestyle{empty} 
%\footnotesize
%\small
\newdimen\digitwidth
\setbox0=\hbox{2}
\digitwidth=\wd0
\catcode`#=\active
\def#{\kern\digitwidth}

\newdimen\ltwidth
\setbox0=\hbox{$<$}
\ltwidth=\wd0
\catcode`@=\active
\def@{\kern\ltwidth}

\newdimen\colonwidth
\setbox0=\hbox{:}
\colonwidth=\wd0
\catcode`!=\active
\def!{\kern\colonwidth}

\begin{deluxetable}{lcccc}
\tablenum{5}
\tablewidth{0pt}
\tablecaption{Ephemerides of FIN 332Aa,Ab \& Ba,Bb}
\tablehead{
\colhead{Epoch} &
\multicolumn{2}{c}{FIN 332Aa,Ab} & 
\multicolumn{2}{c}{FIN 332Ba,Bb} \nl
\colhead{~} &
\colhead{$\theta$} &
\colhead{$\rho$} &
\colhead{$\theta$} &
\colhead{$\rho$} \nl
\colhead{~} &
\colhead{(deg)} &
\colhead{(arcsec)} &
\colhead{(deg)} &
\colhead{(arcsec)} 
}
\startdata
2010.0 & 312.0 & 0.160 & ##6.8 & 0.043 \nl
2010.5 & 311.6 & 0.158 & ##1.2 & 0.048 \nl
2011.0 & 311.2 & 0.155 & 356.6 & 0.053 \nl
2011.5 & 310.8 & 0.151 & 352.8 & 0.058 \nl
2012.0 & 310.3 & 0.147 & 349.6 & 0.063 \nl
2012.5 & 309.9 & 0.143 & 346.9 & 0.068 \nl
2013.0 & 309.4 & 0.139 & 344.5 & 0.072 \nl
2013.5 & 308.8 & 0.134 & 342.4 & 0.077 \nl
2014.0 & 308.3 & 0.128 & 340.5 & 0.081 \nl
2014.5 & 307.6 & 0.122 & 338.8 & 0.085 \nl
\enddata
\end{deluxetable}
%\end{document}

\end{document}